\documentclass[12pt]{article}
\usepackage{graphicx}
\newcommand{\vect}[1]{\mbox{\boldmath $#1$}}   

\begin{document}
\begin{center} {\Large\bf Electron Bernstein waves in spherical tokamak
plasmas with "magnetic wells"
}\\[11pt]
{\large Piliya A.D., Popov A.Yu., Tregubova
E.N.}\\
[6pt] {\it Ioffe Physico-Technical Institute, St.Petersburg,
Russia}\\
[6pt] {e-mail: alex.piliya@mail.ioffe.ru, a.popov@mail.ioffe.ru}\\
\end{center}

\section{Introduction}
Propagation and electron cyclotron resonance (ECR) damping of
electron Bernstein waves (EBWs) in spherical tokamaks (ST) is
usually analyzed assuming that absolute value of the tokamak
magnetic field $|B|$ increases inward the plasma. In this case the
perpendicular index of refraction $n_\perp$ grows as the wave
approaches the ECR layer, where the wave is fully absorbed
regardless of the resonance harmonic number ~\cite{PPT}.
\begin{figure}[h]
\centering
\includegraphics[height=7 cm,bb= 102 382 512 777,clip]{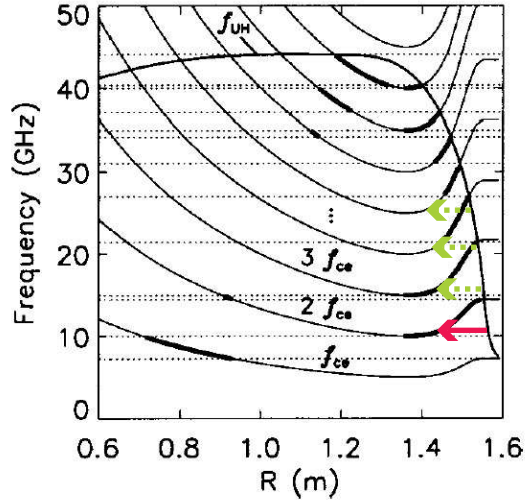}
\caption{Total magnetic field $\vect{B}$ profile in NSTX tokamak
mid-plane ~\cite{NSTXfig}.}
\end{figure}
However, in addition to traditional regimes with monotonously
increasing $|B|$, regimes with "magnetic wells" also occur in STs
~\cite{NSTXfig}. The the magnetic field profile inversion modifies
significantly the whole picture of the wave propagation and
damping. Since the magnetic wells may become quite common with
further improvement of ST performance (fig.1), analysis of such
configurations is of interest for assessment of
EBW plasma heating an CD perspectives.\\
\begin{figure}[h]
\centering
\includegraphics[height=7 cm,bb= 245 460 567 830,clip]{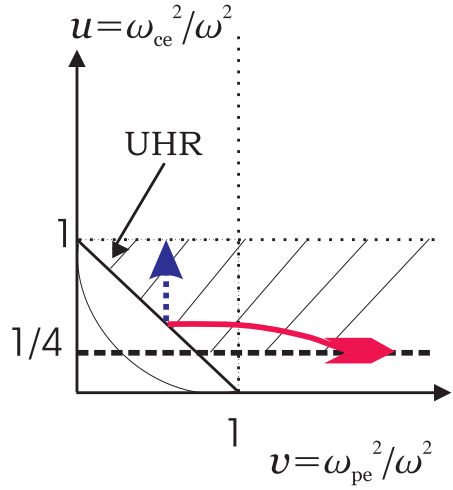}
\caption{CMA diagram. In the case of the conventional tokamak EBWs
can propagate from UHR to the fundamental harmonic (dotted arrow).
In the ST EBWs can propagate to the second EC harmonic (solid
arrow). This case will be considered in this paper.}
\end{figure}
Consider CMA diagram (fig.2) illustrating why this case was not
considered in literature yet. In the conventional tokamaks EBWs
produced via the linear conversion of the incident electromagnetic
waves close to UHR are capable for propagating to the fundamental
harmonic along dotted blue arrow. If the inequality
$d[ln(n_e)]/dx>2\omega_{ce}^2/\omega_{pe}^2d[ln(|B|)]/dx$ is
fulfilled (it's the case of STs) EBWs can propagate along solid
red arrow to the nearest EC harmonic. In this paper we consider
basic features of the EBWs propagation and damping for the second
cyclotron harmonic, which is now the lowest possible resonance
harmonic, in a slab model. This case is illustrated in fig.1 by
solid red arrow.
\section{Bernstein waves in electrostatic approximation}
Assume all plasma parameters depending on the single dimensionless
co-ordinate $x$ scaled in units of $c/\omega$ with $\omega$ being
the wave frequency. Suppose that the magnetic field $\vect{B}$ is
along the $z$ - axis and both $\upsilon=\omega_{pe}^2/\omega^2$
and $q=\omega/\omega_{ce}$ grow inward the plasma.\\
We begin analysis of the EBW behavior assuming validity of the
electrostatic approximation. Then the dispersion relation is
\begin{equation}
\varepsilon\left(n_\perp,n_\parallel,x\right)=0,
\end{equation}
where $\varepsilon\equiv\left(n_i\varepsilon_{ik}n_k\right)/n^2$
is the longitudinal dielectric function, $\varepsilon_{ik}$ - the
plasma dielectric tensor elements, $n_\perp=k_xc/\omega$ and
$n_\parallel=k_\parallel c/\omega$. Since the UHR is known to be
the cutoff for electrostatic EBWs, one expand the dielectric
tensor components $\varepsilon_{ij}, i,j=(x,y,z)$ in power of
$\lambda=k_\perp^2\rho_e^2=\left(n_\perp q\beta\right)^2/2$, where
$\rho_e$ is the electron Larmor radius. As it will be seen,
$n_\perp$ remains within the limits $1\ll n_\perp\leq \beta^{-1}$,
where $\beta=\nu_{te}/c$, $\nu_{te}=\sqrt{2T_e/m_e}$, in the whole
region between the UHR and the $q=2$ resonance, so that
$\lambda\leq 1$ here. For qualitative investigation we keep only
zero- and first-order terms. Then
\begin{equation}
\varepsilon=\varepsilon^{(c)}-\frac{\upsilon\beta^2}{2n^2}\left(3n_\parallel^4+\frac{6q^6-3q^4+q^2}{\left(q^2-1\right)^3}
n_\parallel^2n_\perp^2+\frac{3q^4}{\left(q^2-1\right)\left(q^2-4\right)}n_\perp^4\right),
\end{equation}
where
$\varepsilon^{(c)}=\left(\varepsilon_{xx}^{(c)}n_\perp^2+\varepsilon_{zz}^{(c)}n_\parallel^2\right)/n^2$
and
\begin{equation}
\varepsilon_{xx}^{(c)}=1-\frac{\upsilon q^2}{q^2-1},
\varepsilon_{zz}^{(c)}=1-\upsilon\nonumber
\end{equation}
are the dielectric tensor elements in the cold plasma. The
electron Bernstein waves are produced via the linear conversion of
incident electromagnetic waves with $n_\parallel\leq 1$, while
applicability of the electrostatic approximation requires that
$n_\perp\gg 1$. This permits one to omit terms proportional to
$n_\parallel^2$ and $n_\parallel^4$ in Eq.(2) reducing the
dispersion relation (1) to
\begin{equation}
\varepsilon_{xx}=\varepsilon_{xx}^{(c)}-\frac{1}{2}\upsilon\beta^2\frac{3q^4}{\left(q^2-1\right)\left(q^2-4\right)}n_\perp^2=0.\nonumber
\end{equation}
The solution to this equation
\begin{equation}
n_\perp^2=-\frac{2}{3\beta^2\upsilon}\varepsilon_{xx}^{(c)}\frac{\left(q^2-1\right)\left(4-q^2\right)}{q^4},
\end{equation}
where $\varepsilon_{xx}^{(c)}<0$ and $1<q<2$ shows that in the
electrostatic approximation the EBW is confined between two
cut-offs. One of them is the UHR, for which
$\varepsilon_{xx}^{(c)}(x)=0$, and the other one is the cyclotron
resonance $q=2$. The validity condition for the electrostatic
approximation is $n^2\gg\left|\varepsilon_{ik}\right|$. We
consider here typical for STs high - density plasmas with
$\upsilon\gg 1$, then the validity condition becomes
\begin{equation}
n_\perp^2\gg\upsilon.
\end{equation}
This condition breaks down close to the UHR and ECR. We do not
consider first of these regions because analysis of wave behavior
there is the subject of the mode coupling theory. In the dense
plasma outside immediate UHR vicinity, Eq.(5) becomes
\begin{equation}
n_\perp^2=\frac{2}{3\beta^2}\frac{\left(4-q^2\right)}{q^2}\nonumber
\end{equation}
so that the characteristic value of $n_\perp$ is $\beta^{-1}$. In
the $2^{nd}$ ECR vicinity, where the inequality (6) is broken down
a full - wave treatment is required.
\section{Full-wave equation near $q=2$ resonance}
To obtain a traceable full - wave hot plasma dispersion relation
for EBWs with $n_\perp\sim n_\parallel$ and $n_\perp\ll
\beta^{-1}$ near the $q=2$ resonance we make some simplifications.
Consider first the dielectric tensor elements $\varepsilon_{ik}$.
It is well known that the elements can be presented as an infinite
sums over cyclotron harmonic number $s,-\infty\ll s\ll \infty$
with each term of the sum related to the ECR $q=s$. Since
parameter $\lambda$ is small, we calculate the resonance ($s=2$)
terms of $\varepsilon_{ik}$ up to the first order in $\lambda$
using the zero order (cold plasma) approximation for non -
resonant terms. In this approximation, the elements
$\varepsilon_{xz},\varepsilon_{zx},\varepsilon_{yz},\varepsilon_{zy}$
vanish, $\varepsilon_{xx}=\varepsilon_{yy}=\varepsilon_\perp$,
$\varepsilon_{xy}=-\varepsilon_{yx}=-i g$ and
\begin{eqnarray}
\varepsilon_\perp=\varepsilon_\perp^{(c)}+\frac{1}{2}\frac{\lambda\upsilon}{n_\parallel\beta}Z\left(\xi\right),\,\,
g=g^{(c)}+\frac{1}{2}\frac{\lambda\upsilon}{n_\parallel\beta}Z\left(\xi\right),
\end{eqnarray}
here $g^{(c)}=-\upsilon |q|/\left(q^2-1\right)$, $Z$ is the plasma
dispersion function defined according to ~\cite{Sw} with the
argument $\xi=(q-2)/qn_\parallel\beta$. Consider now the wave
equations
\begin{eqnarray}
\left(n_\parallel^2-\varepsilon_\perp\right)E_x-igE_y-n_\parallel
n_\perp
E_z\nonumber=0\\
igE_x+\left(n^2-\varepsilon_\perp\right)E_y=0\\
-n_\parallel n_\perp
E_x+\left(n_\parallel^2-\varepsilon_{zz}\right)E_z=0\nonumber
\end{eqnarray}
Absolute values of dielectric tensor elements here are of order
$\upsilon\gg 1$ and, therefore $E_z\ll E_x$. Omitting $E_z$ in the
first equation, that means neglecting terms of order
$n_\perp/\upsilon$ and $1/\upsilon$ compared to unity, obtain the
dispersion relation
\begin{eqnarray}
n_\perp^2\varepsilon_\perp-\left(\varepsilon_\perp-g\right)\left
(\varepsilon_\perp+g\right)=0.\nonumber
\end{eqnarray}
Using Eq.(8), we obtain at $\upsilon\gg 1$
\begin{equation}
n_\perp^2\varepsilon_\perp-\frac{\upsilon^2q^2}{q^2-1}\left(1-\frac{n_\perp^2\beta^2(q-1)}{2q}\frac{Z}{n_\parallel\beta}\right)=0.
\end{equation}
Close to UHR, at large $n_\perp$, $n_\perp^2\gg\upsilon$, the
electrostatic approach is valid, the first term in (10) dominates
and we returns to the electrostatic equation (5). In the ECR
vicinity, at small $n_\perp$, $n_\perp\ll\upsilon$, this term can
be omitted and the solution is found explicitly:
\begin{equation}
n_\perp=\frac{2}{\beta}\sqrt{\frac{n_\parallel\beta}{Z\left((q-2)/qn_\parallel\beta\right)}}.
\end{equation}
Thus, in the vicinity of the $q=2$ resonance, where $n_\perp(x)$
goes down with $x$, there is a spatial region
$\upsilon^2>n_\perp^2>\upsilon$ where Eqs.(5) and (11) are valid
simultaneously.\\
The dispersion relation (10) can be also
used in the "relativistic" case $n_\parallel\leq\beta$ if the
"non-relativistic" plasma dispersion function $Z$ is replaced by a
proper relativistic dispersion function.
\section{Wave behavior in the $2^{nd}$ ECR layer}
Close to the $2^{nd}$ ECR layer we consider, for simplicity, as
$x$ dependent only the resonant factor $(q-2)/q$ in the argument
of the $Z$ function, putting $(q-2)/q=x/l$, where $l$ is the
dimensionless characteristic scale-length of the magnetic field
variation. Using the asymptotic expression $Z\sim-1/\xi$, where
$\xi=x/L$ and $L=l\beta n_\parallel$, at $\left|\xi\right|\gg 1$,
one find that at negative $x$ outside the resonance layer Eq.(11)
differs from the electrostatic equation (5) only by a constant
factor  in the right-hand side. Inside the EC layer at
$\left|\xi\right|\sim 1$ the function $Z$ is a complex function
with $|Z|\sim 1$ and $Re(Z)\sim Im{(Z)}$. Solution to Eq.(11) is
also complex with $\left|n_\perp\right|\sim
\left(n_{\parallel}/\beta\right)^{1/2}$, so that the applicability
condition for the WKB theory in the resonance region takes the
form
\begin{equation}
n_\perp L=l n_\parallel^{3/2}\beta^{1/2}\gg 1,
\end{equation}
The estimation for the relativistic case can be obtained from this
equation by putting $n_\parallel\sim\beta$. We assume that the WKB
approximation is applicable in the region outside the resonance
layer where $\left|n_\perp\right|$ grows. Suppose now that
inequality (12) is satisfied and consider solution to Eq.(11)
inside the resonance layer. This solution in a parameter free form
\begin{equation}
N=\pm Z^{-1/2}, N=n_\perp\left(\beta/(4n_\parallel)\right)^{1/2}
\end{equation}
is shown in fig.3 on the plane $\left(Re(N), Im(N)\right)$.
\begin{figure}[h]
\centering
\includegraphics[height=7 cm,bb= 0 20 300 220,clip]{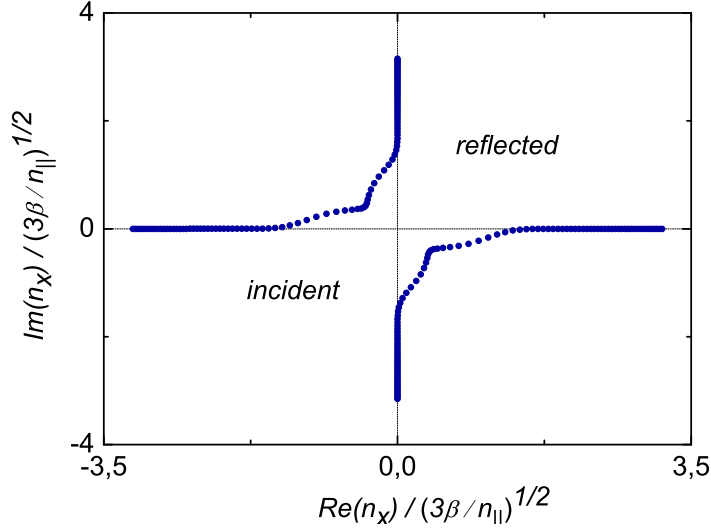}
\caption{Plane $\left(Re(N), Im(N)\right)$. Left branch
corresponds to incident wave, right one - to reflected. In WKB
approximation branches are not intersected. Reflection is the
result of WKB approximation breaking.}
\end{figure}
Brunches located in the upper and lower complex $N$-plane
correspond to the signs $(-)$ and $(+)$ in Eq.(13), respectively.
Consider first the brunch $N= - Z^{-1/2}$. The fragment of the
curve with $Re(N)<0$, $Re(N)\gg Im(N)$ belongs to the region
$x<0$, $|x|\gg L$ when the asymptotic expression for $Z$ can be
used and $n_\perp\rightarrow - \left(4|x|n_\parallel/(\beta
l)\right)^{1/2}$. Since $\left(Im\left(Z\right)\right)^{-1/2}<0$,
the whole curve lies in the upper complex $N$ plane. At $|x|\gg L$
we have $n_\perp\rightarrow i \left(4|x|n_\parallel/(\beta
l)\right)^{1/2}$. The second curve describes transition from
$n_\perp\rightarrow \left(4|x|n_\parallel/(\beta l)\right)^{1/2}$
to $n_\perp\rightarrow -i \left(4|x|n_\parallel/(\beta
l)\right)^{1/2}$ via the lower complex $N$ plane. Note that the
waves under consideration have their phase and group velocities
directed oppositely. Therefore, the incident wave has
asymptotically a negative $n_\perp$ in the propagation region
$x<0$. This wave is described by the $N=-Z^{-1/2}$ dispersion
curve. At large positive $x$ this mode vanishes exponentially. The
second brunch describes the wave reflected from the ECR. This mode
grows exponentially in the non propagation region $|x|\gg L$.Three
important conclusions can be drawn from this analysis. First,
incoming waves incident on the ECR layer from the high-field side
are not converted in the resonance region into outgoing EBWs with
large $n_\perp$ propagating on the low field side of the ECR.
Instead, the incident waves become non-propagating beyond the
resonance layer. Second, two effects are simultaneously
responsible for $Im\left(n_\perp\right)$: the ECR damping and the
wave transition into the non-propagation region. These two
contribution can not be separated. Finally, since the function
$|Z(\xi)^{-1}|$ has no zeros at finite $|\xi|$, two brunches of
the dispersion curves are separated in the whole complex $x$
plane. As a result, reflection from the ECR layer can only be due
to approximate nature of the WKB theory.
\section{Reflection from ECR layer}
We analyze reflection of EBWs from the $q=2$ cyclotron resonance
with the use of the model wave equation obtained from the
dispersion relation (13) by replacement
$n_\perp\rightarrow-id/dx$:
\begin{equation}
U^{''}(x)+n_\perp(x)^2U(x)=0.
\end{equation}
Here $n_\perp^2$ is given by the right - hand side of Eq.(11) and
we assume that function $U$ is related linearly to the wave field
components. More definite interpretation of this function is not
required. In the WKB approximation Eq.(14) has two linearly
independent solutions
\begin{eqnarray}
U_1=\frac{1}{n_\perp^{1/2}}\exp{\left(-i\int_a^x{n_\perp
dx}\right)},\,\,
U_2=\frac{1}{n_\perp^{1/2}}\exp{\left(i\int_a^x{n_\perp
dx}\right)}
\end{eqnarray}
where $a$ is an arbitrary constant. As it has been mentioned
before, $U_1$ and $U_2$ represent incoming and outgoing waves,
respectively. The solution to Eq.(14) describing the ECR damping
and reflection of the wave incident from the high-field side
vanishes at $|x|\gg L$ and has the asymptotic form
\begin{equation}
U=U_1+R U_2
\end{equation}
at $x<0$,$|x|\gg L$. Here $R$ is a constant and the quantity
$|R|^2$ is the energy reflection coefficient. Presenting Eq.(14)
in the form
\begin{equation}
U^{''}(x)+\left(V_0+V_1\right)U(x)=0,
\end{equation}
where $V_0=n_\perp^2-V_1$,
$V_1=\left(n_\perp^{-1/2}\right)^{''}n_\perp^{1/2}$. One can
easily see that functions $U_{1,2}$ are exact solutions to the
equation $U_{1,2}^{''}+V_0U_{1,2}=0$ and $V_1$ is small, of order
$\left(\left|n_\perp L\right|\right)^{-2}$ compared to $V_0$, if
the WKB method applicability condition $\left|n_\perp\right|L\gg
1$ is satisfied. In this case the reflection coefficient $R$ can
be calculating by treating the term in Eq.(17) as a small
perturbation and using $U_1$ as a zero order solution:
\begin{equation}
R=\frac{1}{2}\int_{-\infty}^{\infty}{V_1U_0^2dx}
\end{equation}
or explicitly,
\begin{equation}
R=\frac{1}{\Gamma_0}\int_{-\infty}^\infty{\left(Z^{1/4}\right)^{''}Z^{1/4}\exp{\left(-i\Gamma_0\int_{-\infty}^\xi{Z^{-1/2}\left(\zeta\right)}d\zeta\right)}d\xi},
\end{equation}
where the prime denotes differentiation with respect to $\xi$ and
$\Gamma_0=2l\beta^{1/2}n_\parallel^{3/2}$. For Eq.(19) validity,
$\Gamma_0$ must be large compared to unity. Since the integrand
has no saddle points at finite $|\xi|$, and contribution to the
integral is spread over the whole $|\xi|\leq 1$, the possibility
of obtaining an accurate analytical estimation of the integral
seems rather doubtful. Numerical evaluation shows that
$\left|R\right|^2$ is negligibly small at $\Gamma_0\geq 1$.\\
In the limiting case $\left|k_\perp\right|L<1$ opposite to the WKB
one, the reflection coefficient can be found by presenting
$Z^{-1}(\xi)\simeq -\xi+V$ and treating $V$ as a perturbation.
Substituting these expressions into Eq.(14) and introducing the
new independent variable
$\tau=k_0x\left(\beta^2l/4\right)^{-1/3}$, obtain
\begin{equation}
U^{''}-\left(\tau-\gamma V\left(\tau/\gamma\right)\right)U=0,
\end{equation}
where $\gamma=\Gamma_0^{2/3}$ is a small parameter. In the present
case the reflection coefficient is close to unity. Calculating the
correction to it in the lowest order in $\gamma$ yields
\begin{eqnarray}
|R|\simeq 1-\frac{\Gamma_0^2}{\xi^2},
\end{eqnarray}
where $\xi^{-1}=\pi Ai(0)=1.12$.
\begin{figure}[h]
\centering
\includegraphics[height=8 cm,bb=10 5 290 215 ,clip]{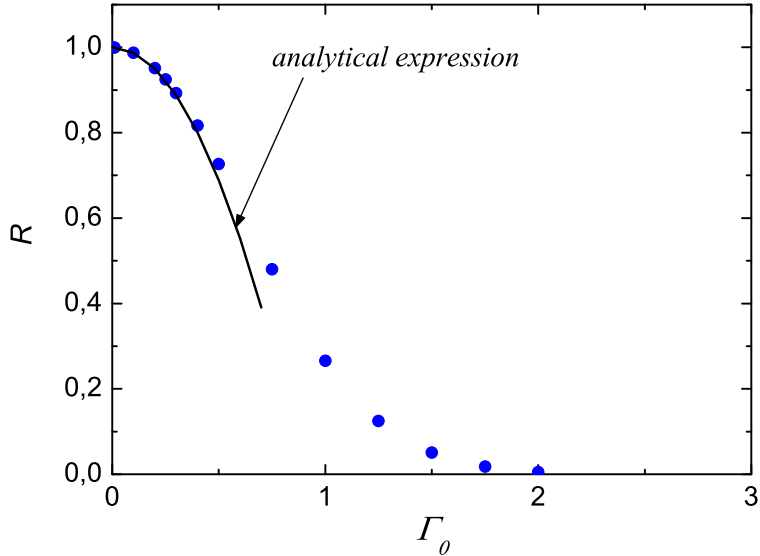}
\caption{Absolute value of the reflection coefficient $|R|$ for
the non-relativistic ($n_\parallel\gg\beta$) case.}
\end{figure}
The comparison of the reflection coefficient, $R$, calculated
numerically and its analytical estimation (21) is presented in
fig.4. The reasonable agreement is demonstrated.
Note in conclusion
that standard ray tracing method using the full wave dispersion
relation or the approximate dispersion relation (11) can be used
for high-field side propagation only if the wave is damped in the
periphery of the ECR, where $Im(n_\perp)\ll Re(n_\perp)$.
Otherwise the wave penetrates into the region of strong cyclotron
damping where the ray tracing method is inapplicable.
\section{Summary}
Existence of EBWs in the region between the UHR and ECR requires
inhomogeneous plasma density.\\
Waves in this region are adequately described by the approximate
full-wave dispersion relation (10). Incoming waves incident on the
ECR layer from the high-field side are not converted in the
resonance region into outgoing EBWs propagating on the low field
side of the ECR. Instead, the incident waves become
non-propagating beyond the resonance layer.\\
Decreasing of the wave amplitude within ECR layer is due to
combine effect of the ECR damping and non-propagation.\\
In the WKB approximation, the waves are fully damped in the ECR
layer.  Reflection from the ECR layer is only due to approximate
nature of the WKB theory.\\
Standard ray tracing method can be used for high-field side
propagation only if the wave is damped in the periphery of the ECR
layer. Otherwise the wave penetrates into the region of strong
cyclotron damping where the ray tracing method is inapplicable.
\\
\\
{\it Acknowledgment:} This work has been supported by RFBR
04-02-16404, 02-02-17683, Scientific School grant 2159.2003.2.
\\
\\

\end{document}